\documentclass[letterpaper]{article} 
\usepackage{aaai24} 
\usepackage{times}  
\usepackage{helvet}  
\usepackage{courier}  
\usepackage[hyphens]{url}  
\usepackage{graphicx} 
\usepackage{natbib}  
\usepackage{caption} 
\usepackage{algorithm}
\usepackage{algorithmic}
\usepackage{graphicx}
\usepackage{times}
\usepackage{helvet}
\usepackage{amsmath}
\usepackage{courier}
\usepackage{xcolor}  
\usepackage{bm}
\usepackage{multirow}
\usepackage{makecell}
\usepackage{graphicx}
\usepackage{newfloat}
\usepackage{listings}

\def\eg{\emph{e.g.}, }

\DeclareCaptionStyle{ruled}{labelfont=normalfont,labelsep=colon,strut=off} 
\floatstyle{ruled}
\newfloat{listing}{tb}{lst}{}
\floatname{listing}{Listing}

\title{Towards Robust Text Retrieval with Progressive Learning}
\author {
    Tong Wu,
    Yulei Qin,
    Enwei Zhang,
    Zihan Xu,
    Yuting Gao,
    Ke Li,
    Xing Sun
}
\affiliations {
    \Large{\textbf{Tencent Youtu Lab.}} \\   
    \{townswu, yuleiqin, miyozhang, ianxxu, yutinggao, tristanli, winfredsun\}@tencent.com
}

\begin{document}

\maketitle

\begin{abstract}
Retrieval augmentation has become an effective solution to empower large language models (LLMs) with external and verified knowledge sources from the database, which overcomes the limitations and hallucinations of LLMs in handling up-to-date and domain-specific information.
However, existing embedding models for text retrieval usually have three non-negligible limitations.
First,
the number and diversity of samples in a batch are too restricted to supervise the modeling of textual nuances at scale.
Second,
the high proportional noise are detrimental to the semantic correctness and consistency of embeddings.
Third,
the equal treatment to easy and difficult samples would cause sub-optimum convergence of embeddings with poorer generalization.
In this paper, we propose the PEG, a progressively learned embeddings for robust text retrieval. Specifically, we increase the training in-batch negative samples to 80,000, and for each query, we extracted five hard negatives. Concurrently, we incorporated a progressive learning mechanism, enabling the model to dynamically modulate its attention to the samples throughout the entire training process. Additionally, PEG is trained on more than 100 million data, encompassing a wide range of domains (\eg{ finance, medicine, and tourism}) and covering various tasks (\eg{ question-answering, machine reading comprehension, and similarity matching}). Extensive experiments conducted on C-MTEB and DuReader demonstrate that PEG surpasses state-of-the-art embeddings in retrieving true positives, highlighting its significant potential for applications in LLMs. Our model is publicly available at \url{https://huggingface.co/TownsWu/PEG.}
\end{abstract}

\section{Introduction}
Information (knowledge) retrieval, a crucial aspect of natural language processing, gains even greater significance in the context of large language models (LLMs)~\cite{ouyang2022training,nakano2021webgpt,openai2023gpt4,touvron2023llama,chowdhery2022palm,workshop2022bloom,zhang2022opt,sun2021ernie}.
The employment of a retrieval model to incorporate external knowledge is essential to enhancing the accuracy and validity of answers generated by LLMs.
Most existing approaches utilize the dense passage retrieval (DPR)~\cite{gao2021condenser,qu2020rocketqa,izacard2021unsupervised,gao2021simcse,ren2021pair,lu2022ernie,khattab2020colbert} for development of retrieval models.
Its steps are composed of the text encoding and text matching, where the encoder of any off-the-shelf language model is used to map queries and a pool of documents into 
representations in the embedding space,
and then similarity between queries and document fragments are measured to match the most relevant candidates.

\begin{figure*}[h]
\centering
\includegraphics[width=0.9\textwidth]{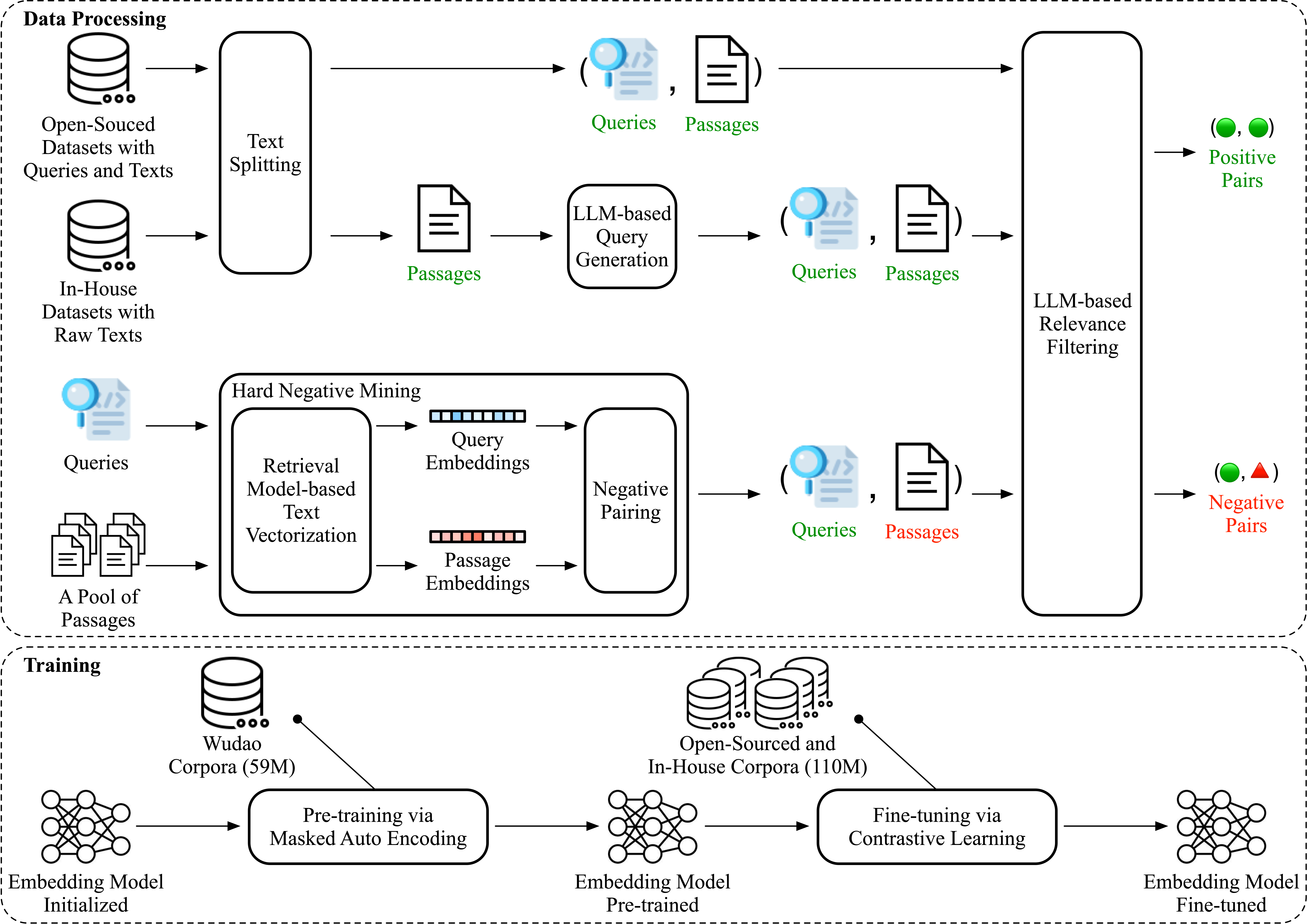}
\caption{
The pipeline of the proposed PEG.
During data processing,
we first split raw texts into short passages and then generate questions from these passages as queries.
Each query is paired with one passage as one positive sample.
In addition,
for each query,
five most similar negatives are retrieved for hard-negative mining.
During training,
we first pre-train the embedding model via masked auto encoding and then perform fine-tuning via contrastive learning. 
}
\label{fig:pipeline}
\end{figure*}

In the field of text encoding, contrastive learning (CL) has emerged as one of the most intuitively effective methods for training embeddings~\cite{gao2021simcse,izacard2021unsupervised,neelakantan2201text,xiao2023c}. This approach aims to minimize the distance between similar, positive sample pairs, while simultaneously maximizing the distance between dissimilar, negative pairs. Given the high cost associated with collecting large-scale labeled corpora, the training process is typically divided into two stages: 1) task-agnostic unsupervised pre-training, and 2) task-specific supervised fine-tuning. During the first stage, method such as SimCSE~\cite{gao2021simcse} employs random augmentation (\eg{dropout}) on the output layer to generate two highly similar yet non-identical counterparts. CL is then performed on these two equivalents as a positive pair, while the remaining samples in a batch are paired with the current sample as negatives for optimizing embeddings. In the second stage, human annotations are used to verify positive and negative pairs. Typically, each query is positively associated with only one passage, while all other passages in a batch are considered negatives.

One challenge associated with contrastive learning based embedding learning is that the representation capacity is closely tied to the quality and quantity of negative samples. A small batch size containing insufficient high-quality, diversified negative samples may not effectively compel the model to discern the subtle differences among highly similar samples, thus impeding its ability to achieve superior discrimination. Consequently, BGE~\cite{xiao2023c} substantially enhances the training batch size, allowing for more than 60,000 negative samples in each batch during the training process. Moreover, it incorporates a hard negative mining approach for offline data processing, in which numerous similar negative samples are sifted using external retrieval models. Although BGE tackles the issue of quantity and diversity of negative samples in text retrieval, it still possesses certain limitations.  First, it relies on a powerful text feature extractor, and if the feature model performs poorly, it may introduce false negatives, resulting in significant noise. Secondly, it assigns equal weight to all negatives, disregarding the varying learning difficulties of easy and hard negatives, which ultimately leads to sub-optimal convergence.

In order to further improve the generalization and robustness of the text retrieval model, we introduce \emph{PEG}, a \textbf{P}rogressively Learned Textual \textbf{E}mbeddin\textbf{G}. First and foremost, we have amassed an extensive collection of over 110 million data, spanning a wide range of fields such as general knowledge, finance, tourism, medicine, and more. This data encompasses a diverse array of downstream tasks, including question-answering (QA) tailored for short text retrieval and machine reading comprehension (MRC) designed to optimize long text retrieval. Secondly, for each query, we have carefully extracted 1 to 5 hard negatives from the dataset. We initially perform an off-line retrieval to obtain the top-5 most similar negatives of each query from the dataset.
Then, we employ a LLM for further cleansing and refinement. If the LLM deems a negative to be highly similar to the query, the negative is then filtered out. Furthermore, by leveraging substantial computational resources, we are capable of accommodating up to 84,000 negative samples within a single batch. And we progressively assign varying weights to different negative samples according to the learning difficulties during different stages of the training process, thereby facilitating the learning procedure.

Extensive experiments conducted on various downstream benchmarks showcase the effectiveness of the proposed PEG, particularly on C-MTEB~\cite{xiao2023c}. We have achieved state-of-the-art (SOTA) performance.

Our main contributions are summarized as follows:

\begin{itemize}

\item We have collected a large-scale retrieval training dataset, consisting of 110 million queries, where each query is paired with one positive sample and five carefully selected hard negatives.

\item We propose the PEG model, which progressively adjusts the weights of samples contributing to the loss within an extremely large batch, based on the difficulty levels of negative samples. 

\item Extensive experiments demonstrate PEG achieves the SOTA on several benchmarks.

\end{itemize}

\section{The proposed PEG}

In this section, we provide a detailed explanation of the proposed PEG.
We begin by introducing the data collection and processing procedure, followed by a discussion on the pre-training and fine-tuning steps (see Fig.~\ref{fig:pipeline}).

\begin{figure*}[htbp]
\centering
\includegraphics[width=\textwidth]{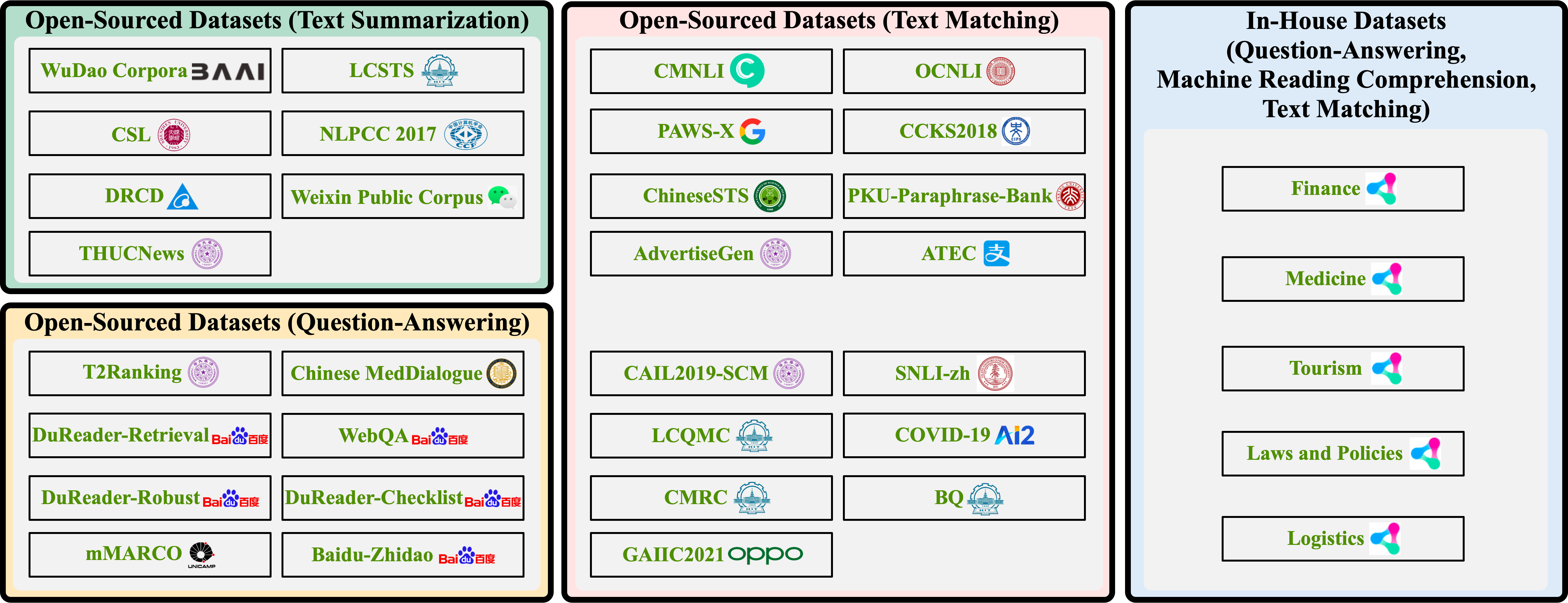}
\caption{Overview of the datasets.
Both open-sourced and in-house datasets are employed across a variety of tasks and domains.
}
\label{fig:dataset}
\end{figure*}

\subsection{Dataset Source}

\paragraph{Pre-training.} We make use of the publicly available Wudao Corpora~\cite{yuan2021wudaocorpora}, which is a huge and high-quality dataset for Chinese language model pre-training.
The data is in the format of title-body, and the total amount of data is 59 million.

\paragraph{Fine-tuning.} We collected 110 million data for fine-tuning (see Fig.~\ref{fig:dataset}).
The vast majority of our data comes from open-sourced datasets while only a small portion of our dataset are privately constructed.
The open-sourced datasets cover a variety of tasks such as text summarization, question answering (QA), and text matching. For the summarization task, we utilize title-passage datasets like Wudao~\cite{yuan2021wudaocorpora},
LCSTS~\cite{hu2015lcsts},
WeiXin Public Corpus\footnote{\url{https://github.com/nonamestreet/weixin_public_corpus}},
CSL~\cite{li2022csl},
NLPCC 2017~\cite{huang2018natural},
DRCD~\cite{shao2018drcd},
and THUCNews~\cite{sunthuctc}. For the QA task,
we utilize datasets like DuReader-Retrieval~\cite{qiu2022dureader_retrieval}, WebQA~\cite{li2016dataset}, T2Ranking~\cite{xie2023t2ranking},
mMARCO~\cite{bonifacio2021mmarco},
Chinese Medical Dialogue~\cite{chen2020meddialog},
BaiDu-Zhidao\footnote{\url{https://github.com/liuhuanyong/MiningZhiDaoQACorpus}},
DuReader-Robust~\cite{tang2020dureader_robust},
and DuReader-Checklist\footnote{\url{https://github.com/baidu/DuReader/tree/master/DuReader-Checklist}}.
For the text matching task, we use CMNLI~\cite{xu2020clue},
OCNLI~\cite{hu2020ocnli},
LCQMC~\cite{liu2018lcqmc},
PAWS-X~\cite{yang2019paws},
CCKS2018~\cite{zhang2019overview},
COVID19~\cite{wang2020cord}, ChineseSTS\footnote{\url{https://github.com/IAdmireu/ChineseSTS}}, CMRC~\cite{cui2019span},
AdvertiseGen\footnote{\url{https://huggingface.co/datasets/shibing624/AdvertiseGen}},
ATEC\footnote{\url{https://github.com/IceFlameWorm/NLP_Datasets/tree/master}},
BQ\footnote{\url{http://icrc.hitsz.edu.cn/info/1037/1162.htm}},
GAIIC2021-OPPO\footnote{\url{https://tianchi.aliyun.com/competition/entrance/531851/introduction}},
CAIL2019-SCM~\cite{xiao2019cail2019},
and PKU-Paraphrase-Bank~\cite{zhang2019pku}.
Our in-house datasets are primarily composed of high-quality books and journals,
covering domains of finance, medicine, tourism, laws and policies, and logistics.

\subsection{Data Processing}

\paragraph{Pre-training.}
We follow BGE~\cite{xiao2023c} to directly use the raw texts in Wudao Corpora~\cite{yuan2021wudaocorpora} for pre-training without additional pre-processing.

\paragraph{Fine-tuning.}
In consideration of the length of articles and chapters in various corpora,
we first split raw texts into shorter passages.
We then make full use of an off-the-shelf LLM to generate questions based on these short passages.
Each question is treated as one query positively associated to the passage,
where the pair $\langle$passage, question$\rangle$ forms a positive sample pair.

At the same time, we allocate five hard negatives to each query.
Specifically, we utilize an open-source retrieval model (\eg{text2vector}) to pinpoint the five most similar samples (excluding the paired positives) for each query as hard negatives.
After retrieving five hard negatives using the retrieval model,
we further employ a LLM to determine the relevance between the query and each hard negative.
If any negative sample is related to the query,
it will be removed to avoid false negatives.

\subsection{Training}

\paragraph{Pre-Training.} Our model is pre-trained on the Wudao corpora, an extensive and high-quality dataset specifically tailored for Chinese language model pre-training. Furthermore, we utilize the MAE-style approach, as presented in RetroMAE~\cite{liu2022retromae}, to train the model effectively. The corrupted text $\hat{X}$ is transformed into its embedding representation, from which the clean text $X$ is reconstructed using a lightweight decoder. The objective of pre-training can be defined as follows: 

\begin{equation}\label{pretraining}
\mathcal{L}_{pt} = \sum_{x\in X} -\log \text{Dec}(x|\bm{e}_{\hat{X}}), \bm{e}_{\hat{X}} \leftarrow \text{Enc}(\hat{X}),
\end{equation}
where the Enc and Dec are respectively abbreviations the encoder and decoder respectively.

\paragraph{Fine-Tuning.}

The pre-trained model is fine-tuned using contrastive learning, which improves the model's capacity to differentiate between text pairs by minimizing the distance between positive sample pairs and maximizing the separation between negative sample pairs. We employ the widely-used InfoNCE loss\cite{he2020momentum} for model optimization:

\begin{equation}
\label{finetuning}
\begin{split}
\mathcal{L}_{ft}=
\sum_{(\bm{e}_{q}, \bm{e}_{p})}-\log& \frac{h(\bm{e}_q, \bm{e}_p)}{h(\bm{e}_q, \bm{e}_p) + \sum_{n}^{N}h(\bm{e}_q, \bm{e}_n)},\\
h(\bm{e}_q, \bm{e}_p)=&\exp(\text{sim}(\bm{e}_q, \bm{e}_p)/\tau),\\
h(\bm{e}_q, \bm{e}_n)=&\exp(\text{sim}(\bm{e}_q, \bm{e}_n)/\tau),
\end{split}
\end{equation}
where $q$ and $p$ represent the indices of a query and its corresponding positive sample,
respectively.
The index of a negative sample is denoted as $n\in N$.
Accordingly,
the embeddings $(\bm{e}_q, \bm{e}_p)$ are positive sample pairs and $(\bm{e}_q, \bm{e}_n)$ are negative ones.
$\tau$ is the temperature hyper-parameter.
We use $\text{sim}(\cdot)$ to represent the similarity measurement (\eg{cosine similarity}) between sample pairs.

One non-negligible disadvantage of the InfoNCE loss above is that it overlooks the difficulty of learning various positive and negative samples.
Negative samples exhibit diverse patterns and the degree of their resemblance to the query indicates how difficult it is for the model to learn to identify their distinction.

Under such circumstance,
each negative pair ought to make an unique contribution to the polishing of embeddings.
We consequently propose the progressive learning mechanism to assign adaptive weights to sample pairs of different levels of learning difficulty.
The proposed approach enables the embedding model to focus on simple samples in the initial stages to first gain the preliminary knowledge on similarity measures between sample pairs.
Then,
it gradually shift the model's attention towards more challenging samples as the training progresses.
Given one mini-batch of $B$ positive pairs and $N$ negative pairs,
our objective is defined as follows:
\begin{equation}
\label{curriculum}
\begin{split}
 \mathcal{L}_{ft} = &\sum_{(\bm{e}_q, \bm{e}_p)}  -w_{q}*\log \frac{h(\bm{e}_q, \bm{e}_p)}{h(\bm{e}_q, \bm{e}_p) + \sum_{n}^{N}g(a_{n}, \bm{e}_q, \bm{e}_n)} \\
&g(a_n, \bm{e}_q, \bm{e}_n)=\exp(a_n\cdot\text{sim}(\bm{e}_q, \bm{e}_n)/\tau),
\end{split}
\end{equation}
where the weight $w_q$ and the scaling factor $a_n$ are respectively defined below:
\begin{equation}\label{w_q}
\begin{aligned}
w_{q} &= 
\begin{cases}
1,  &\text{if}\ \text{sim}(\bm{e}_q, \bm{e}_p)\geq\sigma, \\
\text{sim}(\bm{e}_q, \bm{e}_p) / \sigma &\text{otherwise,}
\end{cases}\\
\sigma&=\frac{1}{B}\sum_{(\bm{e}_q, \bm{e}_p)} \text{sim}(\bm{e}_q, \bm{e}_p) - \beta,\\
\end{aligned}
\end{equation}
\begin{equation}\label{a_n}
\begin{aligned}
a_n = 
\begin{cases}
1,  &\text{if}\ \text{sim}(\bm{e}_q, \bm{e}_p)<\sigma\ \text{or}\\
\ \ &\text{sim}(\bm{e}_q, \bm{e}_n) < \text{sim}(\bm{e}_q, \bm{e}_p),\\
t + \text{sim}(\bm{e}_q, \bm{e}_p), & \text{otherwise},
\end{cases}
\end{aligned}
\end{equation}
where $\sigma$ is a threshold and $\beta$ is its margin.
We measure the similarities of all positive sample pairs within a batch as the normalization basis of the current positive pair.
The hyper-parameter $t$ acts as a bias with respect to the similarity between $\bm{e}_q$ and $\bm{e}_p$.
Compared with the vanilla InfoNCE loss (Eq. \ref{finetuning}),
we intuitively consider that the positive sample pairs whose similarity is below a threshold are potentially false positives and therefore their contribution to the total loss should be weighted according to the batch-wise statistics (\eg{the averaged similarity}).
Besides,
we calibrate the dissimilarities between negative pairs for loss penalty by comparing the similarity between each negative pair and the positive pair.
If one negative sample highly resembles the query,
it is reasonable to believe that such a negative is a hard one and consequently extra emphasis should be put on learning the nuances between the query and this negative.

When it comes to the proper scaling for such calibration,
one naive solution is to set a constant as the bias term $t$.
However,
motivated by the momentum mechanism~\cite{he2020momentum,huang2020curricularface},
we further bring in the batch-wise statistics with a consistent and smooth update policy:
\begin{equation}\label{curriculum_t}
\begin{aligned}
t^{(s)} = \alpha \cdot \frac{1}{B}{\sum_{(\bm{e}_q, \bm{e}_p)} \text{sim}(\bm{e}_q, \bm{e}_p)} + (1-\alpha) \cdot t^{(s-1)},
\end{aligned}
\end{equation}
where $t^{(s)}$ refers to the update of $t$ at the $s$-th step during training and $\alpha$ denotes the momentum coefficient.
Initially,
we set $t^{(0)}$ to 0.

With the progress of training,
the scaling factor would not only reflect the overall similarity distributions across batches but also retain the description of the current positive pair.
Given the proposed progressive learning mechanism,
the optimization of embeddings can greatly benefit from the large-scale contrastive learning to improve their discriminability and robustness against noise.

\section{Experiments}
We conducted experiments on two Chinese text retrieval benchmarks (C-MTEB retrieval and DuReader-Retrieval~\cite{qiu2022dureader_retrieval} datasets) and one Chinese text reranking benchmark (C-MTEB reranking).

\renewcommand{\arraystretch}{2.1}
\begin{table*}[h]
\fontsize{7.4}{7.4}\selectfont
\caption{The quantitive results on C-MTEB retrieval task. We use NDCG@10 as evaluation metric.}
\label{Results on C-MTEB.}
\begin{center}
\setlength{\tabcolsep}{1.1mm}{
\begin{tabular}{|c|c|c|c|c|c|c|c|c|c|}
\hline
{Model} &{T2Retrieval} & {MMarcoRetrieval} & {DuRetrieval} &{CovidRetrieval} & {CmedqaRetrieval}  & {EcomRetrieval} & {MedicalRetrieval} & {VideoRetrieval} & {Avg}           \\ \hline
Text2Vec (base)   &46.55 &43.54  &46.92  &49.82   &16.46  &34.6  &26.9  &39.62  &38.05   \\ \hline
Text2Vec (large)   & 50.52 & 45.96 & 51.87 & 60.48 & 15.53& 37.58 & 30.93         & 42.65 & 41.94  \\ \hline
Text2Vec-bge (large)	& 48.64	& 30.06	& 51.36	& 41.22	& 22.27	& 31.08	& 33.08	& 41.38	& 37.38 \\ \hline
M3E (base)      &72.08  &63.84  &72.59  &66.63   &28.49  &48.41  &39.97   &49.86  &55.23   \\ \hline
M3E (large)     &71.6  &58.25  &72.72 &63.8   &29.41  &45.18  &47.85   &46.48  &54.41   \\ \hline
SimCSE &27.98	&32.52	&36.58	&34.06	&13.71	&14.07	&22.07	&20.4	&25.17 \\ \hline
Contriever	&33.55	&44.37	&38.24	&37.34	&14.53	&35.67	&23.44	&41.3	&33.56 \\ \hline
OpenAI-Ada-002  & 69.14  & 69.86  & 71.17  & 57.21  & 22.36  & 44.49  & 37.92  &  43.85 & 52   \\ \hline
BGE (base) & 83.35  & 79.11  & 86.02  & 72.07  &  41.77 & 63.53  & 56.64  & 73.76 & 69.53   \\ \hline
BGE (large)    & \textbf{84.82} & \textbf{81.28}  &  \textbf{86.94} & 74.06  &  42.4 & \textbf{66.12}  & 59.39  & \textbf{77.19}  & 71.53   \\ \hline
Ours   &  84.67 & 80.63 & 86.36 &  \textbf{81.75} & \textbf{42.72} & 65.95 & \textbf{60.34}  & 76.55 &  \textbf{72.37} \\ \hline
\end{tabular}}
\end{center}
\end{table*}

\subsection{Datasets}

\paragraph{C-MTEB.} The Chinese Massive Text Embedding Benchmark (C-MTEB)~\cite{xiao2023c} is presently the most comprehensive and efficient evaluation benchmark for Chinese semantic embeddings. The construction of C-MTEB can be referred to the corresponding English benchmark MTEB~\cite{muennighoff2022mteb}. 
It encompasses a comprehensive range of 6 evaluation task categories, namely the retrieval, reranking, sentence similarity, reasoning, classification, and clustering.
In total,
it incorporate 31 relevant datasets.
We mainly focus on the

retrieval and reranking tasks.
Especially,
the reranking can also be viewed as another kind of retrieval as it resorts the easily-confused candidates according to their relevance to the query.

The retrieval task predominantly encompass the following datasets: T2Retrieval, MMarcoRetrieval, DuRetrieval, CovidRetrieval, CmedqaRetrieval, EcomRetrieval, MedicalRetrieval, and VideoRetrieval.
Both the EcomRetrieval and VideoRetrieval pertain to sentence-level keyword matching and retrieval,
whereas the rest focus exclusively on query-to-passage retrieval.
For the reranking task,
we use T2Reranking, MmarcoReranking, CMedQAv1, and CMedQAv2.

\paragraph{DuReader-Retrieval.}
DuReader-Retrieval dataset~\cite{qiu2022dureader_retrieval} contains training set, development set, test set with the original paragraph corpus.
It is the first large-scale high-quality Chinese paragraph retrieval dataset based on user search logs under real scenarios.
The queries in the dataset are all real user questions from the Baidu search engine,
and the passages in the dataset are all collected from the retrieved results of Baidu.
We evaluate the performance of our model on the development set,
which 

contains 2,000 query samples and a total of 8.09 million paragraphs.

\subsection{Evaluation Metrics}

For C-MTEB retrieval task, we employ the Normalized Discounted Cumulative Gain (NDCG)@10 as our evaluation metric, with the primary objective of concentrating on the accuracy of ranking within the top 10 recall results. For C-MTEB reranking task, the Mean Average Precision (MAP) score
is used as the main metric. And for DuReader-Retrievl, the evaluation metrics used in this task are the Mean Reciprocal Rank (MRR) and Top-K recall (Recall@K). Specifically, we use the Mean Reciprocal Rank (MRR@10) of the top 10 retrieved paragraphs, the recall of the top 1 retrieved paragraphs (Recall@1), and the recall of the top 50 retrieved paragraphs (Recall@50).

\subsection{Implementation details}

We use BERT~\cite{devlin2018bert}-large model as our basic model architecture, with 24 hidden layers, 16 attention heads, and a hidden size of 1024. We train our model on 32 H800 GPUs. For pre-training, we use AdamW~\cite{loshchilov2017decoupled} optimizer, with an initial learning rate of 2e-5 and a linear decay applied to the learning rate. The batch size per GPU is set at 32, the maximum input sequence length is limited to 512, and the model is trained for 3 epochs. For the fine-tuning phase, we employ the same optimizer and learning rate decay as used during pre-training. The initial learning rate is set to 1e-5, with a batch size of 432 per GPU. The maximum sequence length for the input query and the document are respectively 64 and 256. And the model is trained for 5 epochs. In addition, we refer to BGE to add an instruction in front of each query sample for better retrieval performance.
Out of simplicity,
we empirically set $\alpha=0.5$, $\beta=0.1$, and $\tau=0.01$.

\subsection{Experimental Results}

\paragraph{C-MTEB.}
The results on C-MTEB retrieval task and reranking task are shown in Table \ref{Results on C-MTEB.} and \ref{Results on C-MTEB Reranking.} respectively. In the retrieval task, the newly proposed PEG model attains the SOTA performance, as evidenced by the average NDCG@10 across eight distinct datasets. Notably, the PEG method significantly contributes to the improvement of the CovidRetrieval dataset. As a query-to-paragraph retrieval dataset, CovidRetrieval consists of paragraphs extracted from comprehensive articles rather than short answers to queries. The key information pertinent to the query within extensive texts tends to be more dispersed, thereby increasing the complexity of this dataset. This necessitates the use of high-performing embeddings capable of accurately capturing fine-grained semantics. In the context of the reranking task, PEG continues to demonstrate the SOTA results across all evaluated datasets.
\renewcommand{\arraystretch}{2.1}
\begin{table}[h]
\fontsize{7.0}{7.0}\selectfont
\caption{The quantitive results on C-MTEB reranking task. We use MAP as evaluation metric.}
\label{Results on C-MTEB Reranking.}
\begin{center}
\setlength{\tabcolsep}{1.1mm}{
\begin{tabular}{|c|c|c|c|c|c|}
\hline
{Model} &\makecell[c]{T2-\\Reranking} & \makecell[c]{Mmarco-\\Reranking} & {CMedQAv1} &{CMedQAv2}  & {Avg}          \\ \hline

Text2Vec (base)   &63.93	&12.78	&59.16	&59.73	&48.9   \\ \hline
Text2Vec (large)   &64.82	&12.48	&58.92	&60.41	&49.16  \\ \hline
Text2Vec-bge (large)	&63.51	&9.24	&63.42	&63.57	&49.94 \\ \hline
M3E (base)      &66.08	&17.54	&76.16	&77.29	&57.27   \\ \hline
M3E (large)     &66.08	&18.65	&75.51	&76.27	&59.13   \\ \hline
SimCSE 	&61.34	&12.38	&57.04	&57.72	&47.12 \\ \hline
Contriever	&62.16	&13.57	&49.82	&52.28	&44.46 \\ \hline
BGE (base) &66.49	&28.24	&80.11	&84.78	&64.91   \\ \hline
BGE (large)    &66.2	&26.23	&83.01	&85.01	&65.11   \\ \hline
Ours   &\textbf{68.41}	&\textbf{32.77}	&\textbf{83.19}	&\textbf{85.52}	&\textbf{67.42} \\ \hline
\end{tabular}}
\end{center}
\end{table}

\renewcommand{\arraystretch}{1.9}
\begin{table}[!h]
\fontsize{7.2}{7.2}\selectfont
\caption{The quantative results on Du-Retrieval evaluation set (200,000 documents). We use MRR@10, Recall@1 and Recall@50  as evaluation metric.}
\label{Results on Du-Retrieval.}
\begin{center}
\setlength{\tabcolsep}{1.5mm}{
\begin{tabular}{|c|c|c|c|}
\hline
\multirow{2}{*}{Model} &\multicolumn{3}{c|}{Dureader-Retrieval (200,000 documents)} \\
\cline{2-4} & MRR@10 & Recall@10 & Recall@50
       \\ \hline
Text2Vec (base)   &56.29 &44.70  &89.45     \\ \hline
Text2Vec (large)   & 60.28  & 49.35 & 89.75    \\ \hline
Text2Vec-bge (large)   & 61.88  & 52.60 & 87.10    \\ \hline
M3E (base)      &75.36  &65.3  &96.05        \\ \hline
M3E (large)     &76.95  &67.5  &96.65    \\ \hline
SimCSE &48.26  &38.35  &79.8 \\ \hline
Contriever &50.74  &39.55  &85.6 \\ \hline
BGE (base) & 85.39 & 77.85  & 97.80    \\ \hline
BGE (large) & 87.09 & 80.25  & 98.45  \\ \hline
Ours   & \textbf{89.27} & \textbf{84.10} & \textbf{98.50}    \\ \hline
\end{tabular}}
\end{center}
\end{table}

\renewcommand{\arraystretch}{1.9}
\begin{table}[!h]
\fontsize{7.2}{7.2}\selectfont
\caption{The quantative results on Du-Retrieval evaluation set (8 million documents). We use MRR@10, Recall@1 and Recall@50  as evaluation metric.}
\label{Results on Du-Retrieval (all corpus).}
\begin{center}
\setlength{\tabcolsep}{1.5mm}{
\begin{tabular}{|c|c|c|c|}
\hline
\multirow{2}{*}{Model} &\multicolumn{3}{c|}{Dureader-Retrieval (8 million documents)} \\
\cline{2-4} & MRR@10 & Recall@10 & Recall@50
       \\ \hline

BGE (large) & 44.89 & 32.2  & \textbf{92.6}  \\ \hline
Ours   & \textbf{51.34} & \textbf{39.85} & 91.65    \\ \hline
\end{tabular}}
\end{center}
\end{table}

\paragraph{DuReader-Retrieval.}

The DuReader-Retrieval development set contains 2,000 queries that require our model to pinpoint the most relevant passage from the extensive gallery corpus of over 8 million documents. To conserve computational resources, we have randomly selected a subset of 200,000 documents from the original gallery to create a new smaller gallery. As shown in Table \ref{Results on Du-Retrieval.}, the performance of PEG significantly exceeds that of other models. Compared with the BGE large model, our model achieves an approximate 4\% improvement in Recall@1 and a 2\% increase in MRR@10. It's worth noting that the corpus of DuReader-Retrieval's gallery corpus is 200,000, which is double the average size of the C-MTEB's 100,000. Despite this more challenging gallery, our model still outperforms the SOTA methods, which all demonstrates the robustness of our model. Without losing generalization, we also compared our model with BGE large model on the full 8 million documents gallery. The results are shown in Table \ref{Results on Du-Retrieval (all corpus).}.
We observe a clear advantage over BGE.
Specifically, we achieved $6.45\%$ and $7.65\%$ improvements in MRR@10 and Recall@1, respectively.
This further illustrates that under more complex conditions when the number of gallery surges by orders of magnitude, our model is relatively more robust.

\section{Conclusion}

In this paper,
we propose the PEG for robust text retrieval.
Addressing the limited number and diversity of samples, especially the negatives,
we prepare a large-scale training dataset across a variety of domains and tasks.
We increase the batch size up to 80,000 to enable effective contrastive learning.
Furthermore,
we pay extra attention to hard negative mining and introduce the curriculum strategy by progressively assigning adaptive weights to samples according to their learning difficulty at different training stages.
Extensive experiments confirm that the proposed PEG outperforms the SOTA methods in text retrieval amd reranking.
It provides noise-robust text embeddings,
laying a solid foundation for building retrieval augmented LLM systems.

\bibliography{PEG}

\end{document}